\documentclass[twocolumn,showpacs,showkeys,amsmath,amssymb]{revtex4}

\usepackage{graphicx}
\usepackage{dcolumn}
\usepackage{bm}

\begin{document}

\title{Electronic structure of the $c(4\times2)$ reconstructed Ge(001)
surface}

\author{U.~Schwingenschl\"ogl and C.~Schuster}
\affiliation{Institut f\"ur Physik, Universit\"at Augsburg, 86135 Augsburg,
Germany}

\date{\today}

\begin{abstract}
We investigate the electronic structure of the $c(4\times2)$ reconstructed
Ge(001) surface using band structure calculations based on density
functional theory and the generalized gradient approximation. In
particular, we take into account the details of surface reconstruction by
means of well relaxed crystal structures. The surface electronic states are
identified and the local density of states is compared to recent data from
scanning tunneling spectroscopy. We obtain almost perfect agreement between
theory and experiment for both the occupied and unoccupied states, which
allows us to clarify the interpretation of the experimental data.
\end{abstract}


\maketitle

The electronic properties of semiconductor surfaces attract great
attention since these materials are of special interest for technological
applications. Amongst all, the reconstruction of the Si(001) and
Ge(001) surface has been studied most extensively. For the latter,
the surface electronic structure recently has been investigated by
scanning tunneling spectroscopy \cite{gurlu04}. The experiments
show a distinctly structured local density of states (DOS) close to
the Fermi energy, associated with various surface states, see also
\cite{kubby87,kipp95,kubby96}. Detailed
knowledge about these surface states is highly desirable in order
to understand self-organization of ad-atoms on the Ge(001) surface.

Various effects of self-organization have been reported for both Si
and Ge surfaces. For example, self-assembled metallic chains are
formed by In atoms on Si(111) \cite{hill97} and Au atoms on Si(553)
\cite{ahn05}. Electron motion in Ag films grown on the one-dimensional
In chains is restricted to the chain direction, giving rise to quasi
one-dimensional quantized Ag states \cite{nagamura06}. Ultrathin films
of Ag on Ge(111) are found to be strongly influenced by hybridization
between the Ag and Ge surface states \cite{tang06}. For the Ge(001)
surface, Au growth comes along with a large variety of ordering phenomena
as a function of both coverage and growth temperature \cite{wang0405}.
Adsorption of Pt atoms instead of Au leads to well-ordered nanowire
arrays after high-temperature annealing \cite{gurlu03}. These spontaneously
formed chains are thermodynamically stable, literally defect and kink
free, and have lengths up to hundreds of nanometers. However, the
conduction bands of the Pt chains seem to be strongly modified by the
interaction with the Ge substrate \cite{schafer06}. Furthermore, as the
Pt chains provide a confining potential, quantum mechanical interference
between the Ge(001) surface electrons within the self-organized Pt
arrays results in one-dimensional Ge states with energies resembling
the energy levels of a quantum particle in a well \cite{oncel05}.

\begin{figure}
\includegraphics[width=0.4\textwidth]{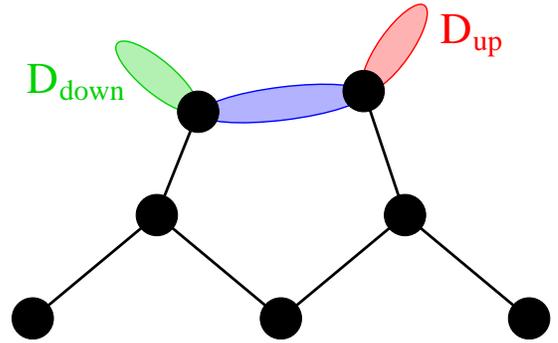}
\caption{(Schematic side-view of a buckled Ge surface
dimer with dangling D$_{\rm down}$ and D$_{\rm up}$ bonds. The dimer bond tilts
slightly out of the surface plane.}
\label{fig1}
\end{figure}

The reconstruction of the Ge(001) surface has been investigated
extensively from both the experimental and the theoretical point of
view \cite{zandvliet03}. In particular, neighbouring surface atoms form
asymmetric dimers on top of a slightly relaxed substrate, saturating
one dangling bond per surface atom, see figure \ref{fig1}. Because the
dimers line up along the $\langle110\rangle$ direction, the Ge(001)
surface is well characterized in terms of dimer rows. Further stabilization
of the surface is reached by a distinct buckling of the dimers, giving
rise to specific reconstruction patterns with respect to the realized
buckling directions. The room temperature $p(2\times1)$ structure
turns into a $c(4\times2)$ structure at low temperatures by means of
an order-disorder phase transition, accompanied by a surface metal-insulator
transition.

Early energy minimization calculations for the reconstructed Si(001)
surface have been performed by Chadi \cite{chadi79}, resulting in the
characteristic dimer geometry as introduced for Ge(001) in the previous
paragraph. Self-consistent electronic structure calculations for both the
Si(001) and Ge(001) surface, using a combination of scattering theory and
the local density functional formalism, have been reported by Kr\"uger
{\it et al.} \cite{kruger86}, see also \cite{pollmann87}. A comparative
study of the surface reconstructions of diamond, Si, and Ge has been
given by Kr\"uger and Pollmann \cite{kruger95}, including a comprehensive
review on previous ab initio calculations. While for diamond a symmetric
dimer configuration is established, asymmetric ordering leads to an
energy gain of about 0.1\,eV per dimer in the case of Si and Ge. The
dimer formation on adsorption of In on Ge(001) has been analyzed by
\c{C}akmak and Srivastava \cite{cakmak04} via first principles total
energy calculations. Their findings stress the importance of the details
of the surface electronic structure for the adsorption mechanism.
Expectedly, the same is true for the adsorption of other atoms or
molecules, as alkali metals \cite{xiao06} or GeH$_4$ \cite{cocoletzi05},
for instance.

Experimental data by Gurlu, Zandvliet, and Poelsema \cite{gurlu04}
indicate that the very details of the surface reconstruction have
serious effects on the Ge(001) surface states. Well relaxed surfaces
therefore are a necessary prerequisite for an adequate theoretical
treatment of the surface electronic structure, which is the aim of
the present paper. To be specific, we calculate the local {\bf k}-integrated DOS for the
surface atoms in order to identify the surface states. Symmetry
analysis then allows us to interrelate these states with particular
bonds of the surface dimers, and to explain the electronic structure
data obtained by Gurlu, Zandvliet, and Poelsema via scanning tunneling
spectroscopy \cite{gurlu04}. Previous theoretical studies failed to
resolve the fine structure of the surface DOS.

\begin{figure}
\includegraphics[width=0.43\textwidth,clip]{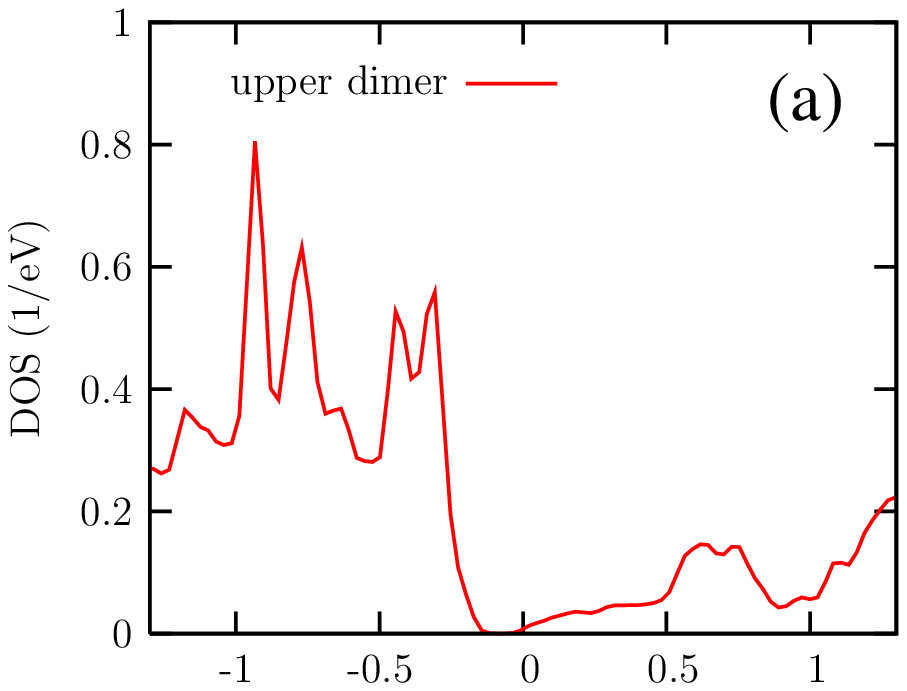}\\
\includegraphics[width=0.43\textwidth,clip]{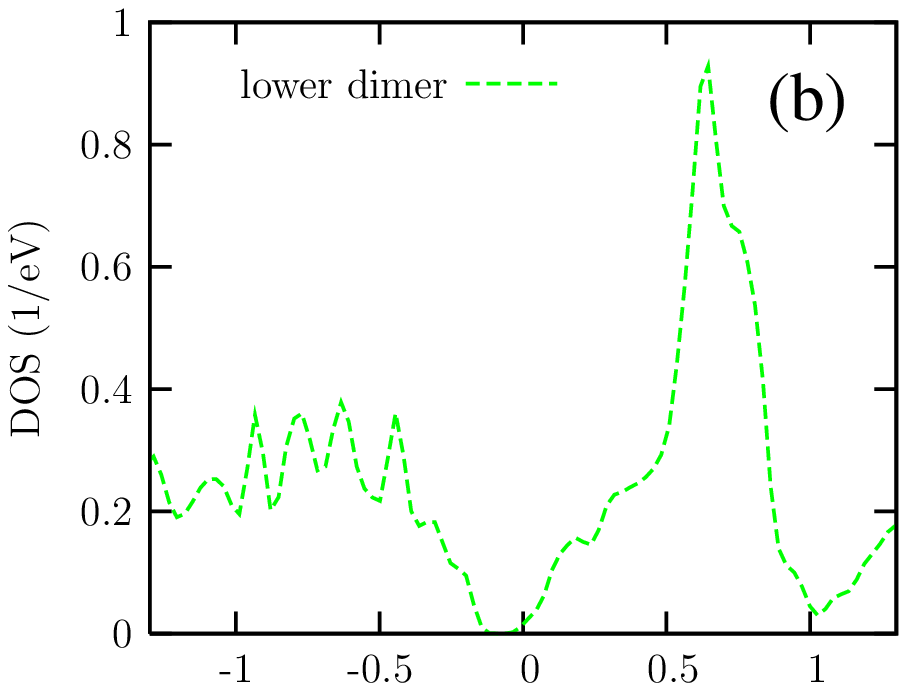}\\
\includegraphics[width=0.43\textwidth,clip]{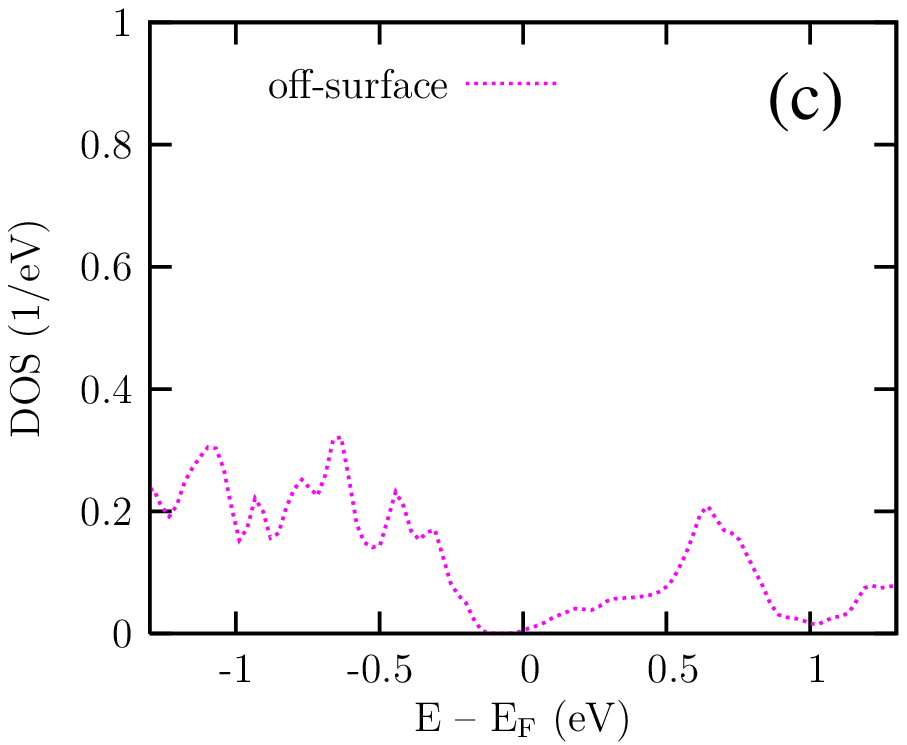}
\caption{Local Ge DOS for (a) the upper dimer site,
(b) the lower dimer site, and (c) an off-surface site.}
\label{fig2}
\end{figure}

Our first principles calculations rely on a supercell of the cubic
Ge unit cell comprising one $c(4\times2)$ reconstructed surface array
and extending two cubic unit cells perdendicular to the surface. With
respect to the parent diamond lattice, our tetragonal supercell
thus is given by the lattice vectors $(4,-4,0)$, $(2,2,0)$, and
$(0,0,2)$. As a consequence, it contains 64 Ge sites in 24 crystallographically
inequivalent classes. For the structural optimization and the
calculation of the electronic band structure, we apply the generalized
gradient approximation within density functional theory, as implemented
in the WIEN2k program package \cite{wien2k}. This is a famous
full-potential linearized augmented plane wave code, having shown
great capability in dealing with structural relaxation at surfaces
and interfaces \cite{uscs}. The x-ray
diffraction data of Ferrer {\it et al.} \cite{ferrer95} and the first
principles data of Yoshimoto {\it et al.} \cite{yoshimoto00} give
rise to a reasonable starting point for the structural optimization.
Convergence of the relaxation is assumed when the surface forces have
decayed. To obtain reliable results, the charge density is represented
by some 230{,}000 plane waves in our calculations and a {\bf k}-mesh
with 40 points in the irreducible wedge of the Brillouin zone is applied.
While Ge $3d$ orbitals are treated as semi-core states, the valence
states consist of Ge $4s$ and $4p$ orbitals. We have checked our band
structure data for convergence with respect to the thickness of
the Ge slab. In particular, the interior of the slab resembles the
bulk Ge DOS.

Figure \ref{fig2} shows the local {\bf k}-integrated Ge DOS as resulting
from our band structure calculations, for three characteristic atomic
sites. The gross features agree well with recent findings of
Stankiewicz and Jurczyszyn \cite{stankiewicz07}. Panels (a) and (b) of
figure \ref{fig2} refer to the upper and lower dimer site, respectively.
While the upper dimer site is shifted off the surface due to the
buckling of the surface dimer, the lower dimer site approaches the
surface. For comparison, panel (c) displays the local Ge DOS for an
off-surface site in the third Ge layer. All three curves in figure
\ref{fig2} show distinct densities of states both below and above the
Fermi energy. The two regions are seperated by an energy gap of about
0.15\,eV, which is considerably less than the experimental gap of
0.7\,eV for bulk Ge. It is likewise less than the values measured
for the Ge(001) surface, amounting to 0.3 -- 0.9\,eV \cite{gurlu04,nakatsuji05}.
A discrepancy between the experimental and theoretical band gap is a
well-known phenomenon, tracing back to the approximations entering
the band calculation. It has to be taken into account when DOS curves
are compared, but does not affect our further conclusions.

To prepare for the analysis of the band structure data, we first
address the surface states expected for the realized reconstruction
pattern. Due to the dimer formation, $\sigma$-type bonding and antibonding
bands should be formed. In addition, each Ge atom is left with one free
dangling bond. For the buckled dimer, the dangling bond of the upward
buckled atom gives rise to the D$_{\rm up}$ surface band, whereas the
dangling bond of the downward buckled atom forms the D$_{\rm down}$ band.
Due to charge transfer from the lower to the upper dimer site, the
energy should be higher for the D$_{\rm down}$ than for the D$_{\rm up}$
state \cite{zandvliet03}.

Comparing the Ge DOS for the upper and lower dimer site, see panels (a)
and (b) of figure \ref{fig2}, we find contributions due to the atomic
orbitals of the upper atom mainly below the Fermi energy. In contrast,
states tracing back to the lower atom dominate above the Fermi energy.
We first address the minority contributions to both the occupied and
unoccupied states by relating them to the DOS of the off-surface Ge
site given in figure \ref{fig2} (c). We observe almost perfect agreement,
even quantitatively, of the lower dimer site DOS and the off-surface
DOS at energies below the Fermi level. The same is true for the upper
dimer site DOS and the off-surface DOS at energies above the Fermi
level. We therefore conclude that the minority contributions to the
occupied and unoccupied bands result from bulk-like states, and are
not characteristic for the Ge surface. Surface states lead to additional
spectral weight, as observed for the upper/lower dimer site below/above
the Fermi energy.

For the upper dimer site, the occupied states show a six peak
structure in figure \ref{fig2} (a), which likewise is visible for the
lower dimer site in panel (b) and the off-surface site in panel
(c). Peaks one (at $-1.2$\,eV) and four (shoulder at $-0.7$\,eV)
are related to bulk-like states, as revealed by the off-surface DOS.
Surface states correspondingly are located at about $-0.9$\,eV, $-0.8$\,eV,
$-0.4$\,eV, and $-0.3$\,eV. For analyzing the unoccupied states, we
turn to the lower dimer site in figure \ref{fig2} (b), showing a
pronounced peak in the energy range from the Fermi level up to 1\,eV.
However, on closer inspection this peak again consists of four surface
states, located near 0.2\,eV, 0.4\,eV, 0.6\,eV, and 0.7\,eV. While
the third state dominates the lower valence band, the other states
contribute less spectral weight, therefore giving rise to distinct
shoulders.

\begin{figure}
\includegraphics[width=0.43\textwidth,clip]{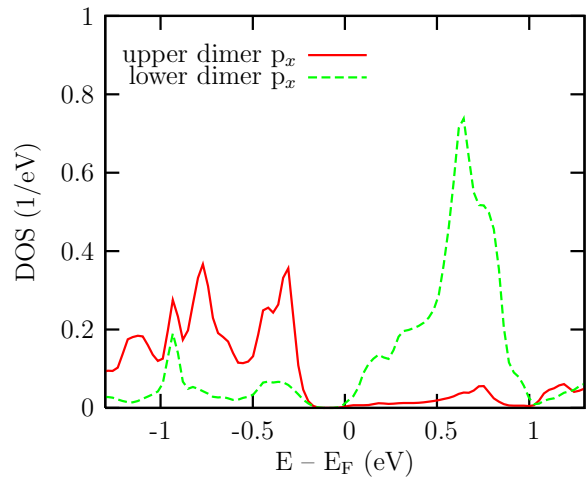}
\caption{Local Ge $4p_x$ DOS for the upper (solid line)
and lower (dashed line) dimer site. The $x$-axis is oriented perpendicular
to the surface.}
\label{fig3}
\end{figure}

Next we investigate the origin of the various surface states. For this
purpose, we decompose the Ge $4p$ DOS into its symmetry components.
Contributions of $p_y$ and $p_z$ symmetry are found to be negligible
in the vicinity of the Fermi energy, but almost all states have $p_x$
symmetry. Since the surface plane is the $yz$-plane in our coordinates,
this implies that an orientation perpendicular to the surface is
typical for all Ge surface states. Corresponding $4p_x$ DOS curves are
given in figure \ref{fig3} for the upper as well as the lower dimer site.
There are only two surface states revealing significant contributions
from both dimer atoms: the energetically lowest state at $-0.9$\,eV
and the energetically highest state at 0.7\,eV. As a consequence, we
attribute these peaks to the $\sigma$-type bonding and antibonding
dimer bands, in correspondence with previous theoretical findings
\cite{pollmann87}. The remaining six surface states purely belong to
either the upper or the lower dimer site. Therefore, the higher occupied
and lower unoccupied surface states are clearly due to the dangling
bonds of the upward and downward buckled Ge atom, respectively.

Turning to the comparison with the experimental DOS obtained by Gurlu,
Zandvliet, and Poelsema \cite[Figure 3]{gurlu04}, there is agreement
as concerns the gross structure of the surface DOS. However, the
experiment does not fully resolve the fine structure. In particular,
the experimental structure at $-0.5$\,eV probably consists of two
peaks, as indicated by a small shoulder on the low energy side,
according to our peaks at $-0.4$\,eV and $-0.3$\,eV. Moreover, our
DOS peaks at $-0.9$\,eV and $-0.8$\,eV are present in the experiment at
slightly different energies. This may be due to the underestimation
of the band gap in our calculation or due to the uncertainties of the
experimental DOS coming along with the numerical evaluation of the
differential conductivity. For the unoccupied bands, differences between
theory and experiment appear to be larger at first glance. However,
our four surface states can be identified in the experimental DOS,
only their relative weights differ. The first peak at 0.2\,eV seems
to correspond to a tiny shoulder right above the Fermi level, which
has not been discussed by Gurlu, Zandvliet, and Poelsema. The peaks
at 0.4\,eV, 0.6\,eV, and 0.7\,eV on the other hand perfectly
agree with experiment.

We finally comment on the interpretation of the experimental surface
DOS. As concerns the $\sigma$-type bonding and antibonding dimer bands
our findings are fully in line with the conclusions of Gurlu, Zandvliet,
and Poelsema. The same is true for the highest occupied and lowest
unoccupied surface state, except that we find a two peak
structure. For the remaining surface bands (at $-0.8$\,eV and 0.6\,eV
in our data) the interpretation of the experiment is difficult. The
authors therefore have to speculate about the origin of these states,
particularly due to a lack of state-of-the-art electronic structure
data for comparison \cite{gurlu04}. Our results clearly show that all occupied surface
states, except for the $\sigma$-type dimer states, originate from the
dangling D$_{\rm up}$ bonds, therefore giving rise to almost pure
D$_{\rm up}$ surface bands. Similarly, the three lowest unoccupied
surface states trace back to the dangling D$_{\rm down}$ bonds, forming
likewise pure D$_{\rm down}$ surface bands.

In conclusion, we have investigated the surface electronic structure
of the $c(4\times2)$ reconstructed Ge(001) surface by means of band
structure calculations within density functional theory. Taking into
account the details of the structural relaxation, we have discussed
the surface density of states and established comprehensive insight
into the various surface states. In particular, it is possible to
attribute each state to either the Ge dimer bonds or the dangling
up/down bonds of the surface atoms. Our band structure results agree
well with the local density of states obtained by spatially resolved
scanning tunneling spectroscopy. As a consequence, we are able to
clarify the interpretation of the experimental data for the $c(4\times2)$
reconstructed Ge(001) surface. We expect that similar studies can
shed new light on related semiconductor surfaces, like Si(001) or GaAs(001).

\subsection*{Acknowledgement}
We thank U.\ Eckern and V.\ Eyert for helpfull discussions, and
the Deutsche Forschungsgemeinschaft for financial support (SFB 484).

\end{document}